# Atomic-Scale Insights into Damage Produced by Swift Heavy Ions in Polyethylene

P. Babaev[1,*], F. Akhmetov[2], S. Gorbunov[1], N. Medvedev[3,4], R. Rymzhanov [5,6], R. Voronkov[1], A.E. Volkov[1]

[1]*P.N. Lebedev Physical Institute of the Russian Academy of Sciences, Leninskij pr., 53,119991 Moscow, Russia*

[2]*Industrial Focus Group XUV Optics, MESA+ Institute for Nanotechnology, University of Twente, Drienerlolaan 5, 7522 NB Enschede, The Netherlands*

[3]*Institute of Physics, Czech Academy of Sciences, Na Slovance 1999/2, 182 21 Prague 8, Czech Republic*

[4]*Institute of Plasma Physics, Czech Academy of Sciences, Za Slovankou 3, 182 00 Prague 8, Czech Republic*

[5]*Flerov Laboratory of Nuclear Research, Joint Institute for Nuclear Research, Dubna, Russia*

[6]*Institute of Nuclear Physics, Almaty, Kazakhstan*

[*] Corresponding author: babaevpa@lebedev.ru

## Abstract

We describe the formation of swift heavy ion tracks in polyethylene (PE) by combining the Monte Carlo code TREKIS, which models electronic excitation in nanometric proximity of the ion trajectory, with the molecular dynamics simulating a response of the atomic system to the perturbation. The model predicts circular tracks in amorphous PE but elliptical ones in crystalline PE caused by preferential propagation of excitation along polymer chains during the cooling stage. The obtained track sizes and shapes agree well with the high-resolution microscopy of tracks in PE. The velocity effect in PE is shown: the track parameters differ for ions with the same energy losses but different velocities.

## I. Introduction

Swift heavy ions (SHI, $E > 1$ MeV/a.m.u., $M \geq 6$ a.m.u.) lose their energy mainly via excitation of the electronic system of a target in the nanometer proximity of their trajectories [1–3]. The passage of an SHI is characterized by a high energy release followed by drastic and fast (~100 ps) structure transformations in a target leading to the formation of ion tracks with extremely high length/radius ratio (~100 μm/10 nm) [3,4].

SHI irradiation triggers effects changing the physical and chemical properties of organic polymers: primary bond cleavage leads to chain scissions and cross-linking; low mass fragments (e. g. $H_2$, $C_xH_y$, $C_xO_z$ in widely used polymers) tend to leave the track reducing its density; unsaturated bonds, free radicals and small molecules such as alcohols and carboxylic acids are formed [5,6]. Excited long organic molecular chains favor the formation of isomers and cross-linking of polymer chains, resulting in a wide variety of polymer structures. Due to the low thermal conductivity of polymers, cooling of the initially heated region with a diameter ≤ 10 nm takes over 100 ps [2], which may induce effects that are absent in inorganic materials.

Technological advances [1] motivate the development of various experimental methods to study damage in latent tracks in polymers. Transmission electron microscopy enables observation of the size and shape of SHI tracks [6,7]. Spectroscopic techniques such as infrared, ultraviolet, visible, and Raman spectroscopy investigate physical and chemical changes in the irradiated polymers [8,9]. Recently, diffraction techniques have been used to investigate the velocity effect in polyethylene terephthalate: dependence of the damaged track size on the SHI velocity at the same linear energy transfer (LET or $S_e$) to the electronic system [10].

Analytical models can describe track parameters in polymers by fitting them to the experimental data. It is possible to determine the profile of changes in the electron density in a track and its characteristic dimensions by fitting SAXS curves [10]. Describing chemical species in a cooled track, a simple model in Ref. [11] translates the radial distribution of the released energy into the radial concentrations of transient types of low-mass fragments, and further into radial concentrations of the final fragments, cross-linked structures, and intact macromolecules.

Modelling is also widely used in investigations of the SHI effects in polymers. Monte Carlo (MC) simulation of a response of the electronic system in polymeric material to the excitation by an SHI impact was reported in Ref. [12]. The approach describes the spatial distribution of ionization events in a track, reproducing inhomogeneity of the radial ionization density and generation of a large number of low-energy electrons. Refs. [13–15] used the scattering cross sections extracted from the experimental optical data for calculations of the electronic inelastic mean free paths, the stopping power, and the radial distributions of the energy deposited in a material by secondary electrons in some polymers.

Progress in the development of the reactive force fields allowed the tracking of atomic trajectories and bond rearrangements in polymers with atomic precision [16–18]. Refs. [19–21] combined the thermal spike model with coarse-grained molecular dynamics (MD), treating monomers as separate simulation blocks (the united-atom approach). This approach allowed to estimate crater formation and sputtering from the surface of polymers irradiated with SHIs. Alternatively, a detailed approach in Ref. [22] models polymers irradiated with SHIs combining the thermal spike model adapted to treat polymers with the modified charge-implicit ReaxFF reactive force field in molecular dynamics, treating each atom as a distinct simulation block (the all-atom approach). The approach reproduced the track structure and revealed chemical transformations during track formation. Computational models describing the etching of SHI tracks are also being actively developed [3,23,24].

The application of hybrid approaches is based on the fundamental separation of the track formation kinetics into a sequence of well-separated stages allowing to construct multiscale models for its description [3,25]. An SHI passes the interatomic distance within $\sim 10^{-3}$ fs, causing extreme electronic excitation. Relaxation of the excitation, accompanied by energy transfer to the lattice, lasts ~50-100 fs. It is followed by the kinetics of the heated atomic ensemble up to track cooling by about ~ 1 ns after the ion passage.

Recently, an approach quantitatively describing all coupled stages of SHI track formation in inorganic dielectrics was developed [26–28]. It is based on the application of MC code TREKIS-3 modeling excitation of the electronic and atomic systems followed by molecular dynamics (LAMMPS) tracing atomic trajectories [29]. The scattering cross sections used in the MC module account for collective responses of the electronic and atomic systems to the excitation in the framework of the loss function (the imaginary part of the inverse complex dielectric function (CDF)) formalism [30,31].

To describe SHI track formation in PE, we applied the same [26–28] methodology, using in MD simulations the reactive force field AIREBO-M for hydrocarbons [18]. On the example of polyethylene, we demonstrate that this approach is a reliable tool for modeling SHI tracks in polymers, which reproduces the track sizes detected in the experiments without *a posteriori* fitting parameters. The obtained spatial distributions of broken bonds and small-mass fragments can be used in models of chemical activation of SHI-irradiated polymers and their etching. We also pointed out the velocity effect: different PE damage by ions with the same LET but different energies.

## II.    Model

We use the combined approach consisting of Monte Carlo modeling the response of the electronic system to an SHI passage and molecular dynamics tracing the atomic system evolution.

The asymptotic trajectory event-by-event MC code TREKIS-3 describes the excitation of the electronic ensemble and calculates the energy transferred to the atomic system [32–34]. TREKIS-3 models the following processes: 1) SHI passage through a target, resulting in the creation of primary electrons and holes; 2) scattering of primary and all secondary electrons on target atoms and electrons; 3) Auger decays of core holes, resulting in the generation of secondary electrons; 4) radiative decays of the core holes with subsequent photon emission and photoabsorption exciting new electrons and holes; 5) transport of valence holes and their interaction with atoms of the target [30,31].

Within the first-order Born approximation, the inelastic scattering cross section of an incident particle (an SHI, an electron, or a valence hole) is expressed in terms of the loss function that takes into account the collective response of a target to excitation [35–37]. The loss function approximated with a set of oscillatory functions can be reconstructed from the optical coefficients using Ritchie and Howie's algorithm [36]. We use the optical coefficients for polycrystalline PE from Ref.[30], because no data were found for crystalline and amorphous PE separately. Both the used coefficients of the loss function and its graph for PE can be found in [30] (Table A7 and Figure A6, respectively) and in [38]. However, the application of the phonon part of the loss function restored from the optical coefficients provides unrealistically large elastic mean free paths of electrons (up to ~1 μm). Thus, we describe the elastic scattering of electrons and valence holes with the Rutherford cross sections with the modified Molier screening parameter [39].

The MC procedure is iterated $10^3$ times to achieve reliable statistics. We obtain the temporal dependencies of the radial distributions of the electron density and energy, the distribution of holes in the valence band and various atomic shells, and the energy transferred to the atomic system of the target. As in Refs. [40,41], in addition to the energy received by atoms due to the scattering of electrons and valence holes, we assume an instantaneous deposition of the potential energy of valence holes (electron-hole pairs) to the lattice at 100 fs after the SHI passage. This energy transfer approximates the effects of atomic acceleration caused by a transient nonthermal modification of the interatomic potential [42,43].

The calculated radial distribution of the energy transferred to the atomic system of the target is then used as input data for the classical MD code LAMMPS [44], which describes atomic response causing damage in the proximity of the projectile trajectory. We set the initial velocities of atoms in coaxial cylindrical layers using the distribution of transferred energy calculated with the MC assuming Gaussian-like dispersion of the kinetic energy and the uniform distribution of the atomic momenta within each layer [40,41]. Interactions between atoms in PE are calculated using the AIREBO-M force field [18].

Although experimental samples usually have a polycrystalline or mixed polycrystalline-amorphous structure, we use fully crystalline and fully amorphous supercells to demonstrate damage characteristics in them.

In the MD simulation, the orthorhombic crystalline supercell of $197 \times 197 \times 50$ Å$^3$ with 259200 atoms is used, with the unit cell parameters taken from Ref. [45]. The usual width of the PE crystalline grain (so-called *lamella*) is of about 200 Å [46], therefore, in the simulations we can see damage across its entire width. Each polymer chain in the crystalline supercell contains 156 (CH$_2$) polymer units (468 atoms). The crystalline sample in our simulation was prepared using the Moltemplate tool [47] and has a density of about 1 g/cm$^3$.

For simulations of the amorphous PE, the supercell has the dimensions of $204 \times 190 \times 50$ Å$^3$ and contains 207168 atoms. Chains in the amorphous supercell have 1,328 atoms (442 carbons and 886 hydrogens) built from CH$_2$ units and have CH$_3$ tags at their ends, with the unit cell parameters from Ref. [45]. The amorphous sample, prepared using the EMC tool [48], has a density of about 0.82 g/cm$^3$.

Both, crystalline and amorphous simulated densities are close to the experimental ones from Ref. [49] (1.003 g/cm$^3$ and 0.85 g/cm$^3$ respectively). Examples of supercell structures are shown in Fig.1.

The periodic boundary conditions are used in all directions. The boundaries of the supercell along the directions perpendicular to the SHI trajectory are cooled by the Berendsen thermostat to 300 K with a characteristic time of 0.1 ps [50]. The supercell is simulated until its average temperature drops below 350 K when no structural changes are expected thereafter. Refs [30,31] contain more details describing the calculations. The results of the simulations are visualized with the help of OVITO software [51].

The selection of the U, Xe, and Zn ions for the simulations is based on the availability of experimental data for comparison [52].

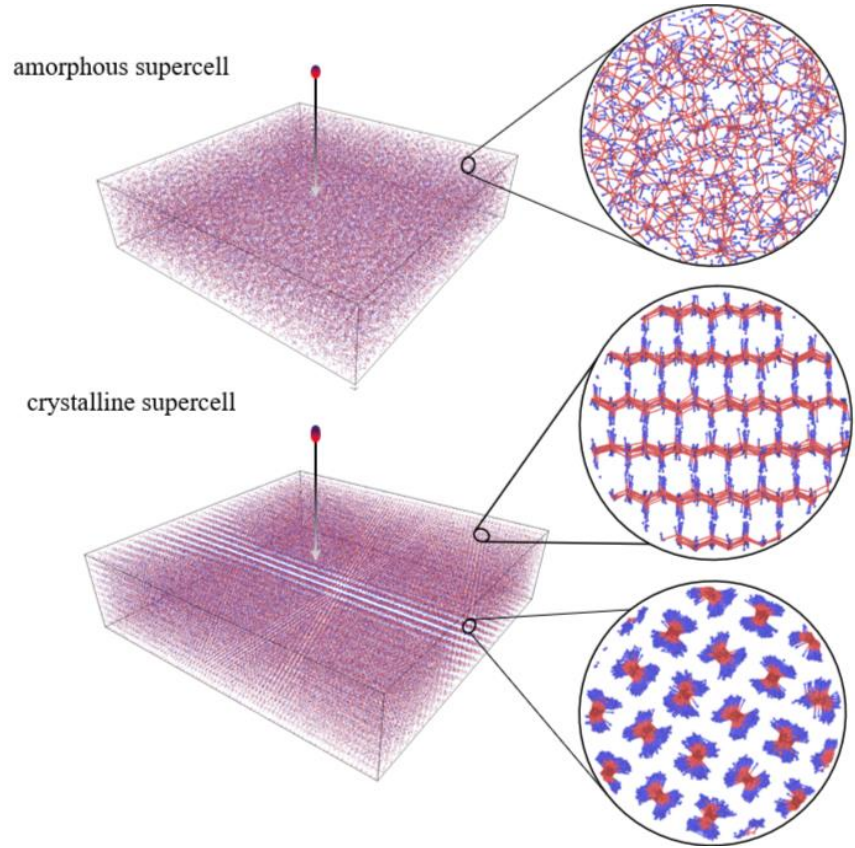

**Figure 1.** Crystalline **(top)** and amorphous **(bottom)** supercells with their magnified structures. Red balls represent carbon atoms, and blue balls are hydrogen.

## III. Results and Discussion

### III.1. Structure changes: amorphous vs. crystalline PE

Figs. 2 and 3 illustrate the snapshots of the supercell before and after 700 MeV U and Xe 1470 MeV ion impacts in amorphous and crystalline PE accompanied by the partial radial pair distribution functions (RPDF), consequently. The changes in the C-C RPDF characterize the cleavage of polymer chains and the appearance of new structures; the C-H RPDF shows the detachment of hydrogen atoms from parent carbon chains; the H-H RPDF shows an appearance of molecular hydrogen (0.75 Å peak). Areas of damaged and less dense material can be seen in the supercell snapshots.

The decrease of the major peaks in C-C, C-H, and H-H RPDFs and the appearance of the minor peaks in C-H and H-H RPDFs indicate severely damaged structures with bond rearrangements and formations of new molecular species. No voids were observed in the track in the simulation, which agrees with the experiment [52].

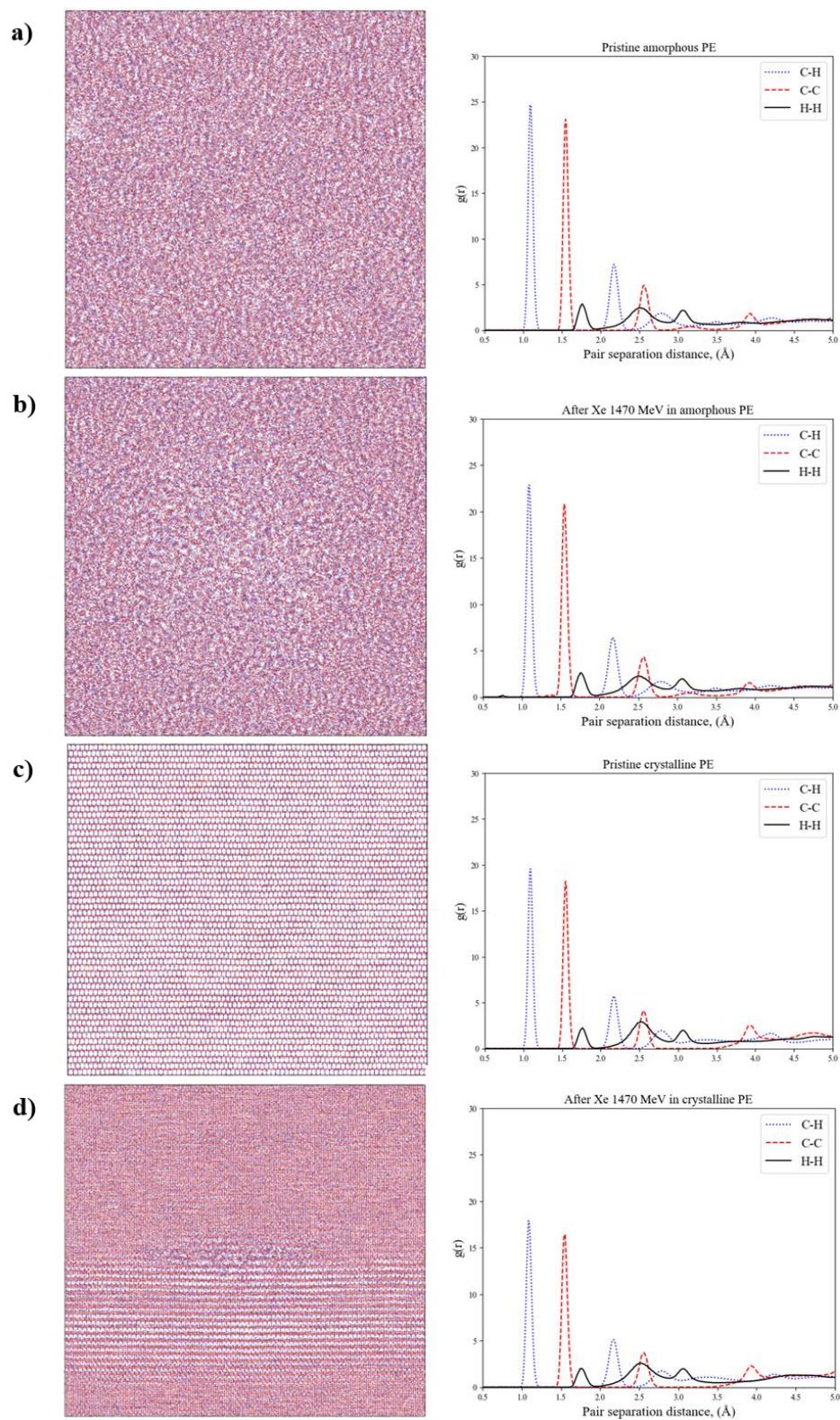

**Figure 2.** MD supercell snapshots (left panels) and RPDFs (right panels) prepared by OVITO [51]: (a) pristine amorphous cell of PE; (b) amorphous one irradiated with Xe

1470 MeV; (c) pristine crystalline PE cell; (d) crystalline one irradiated with Xe 1470 MeV.

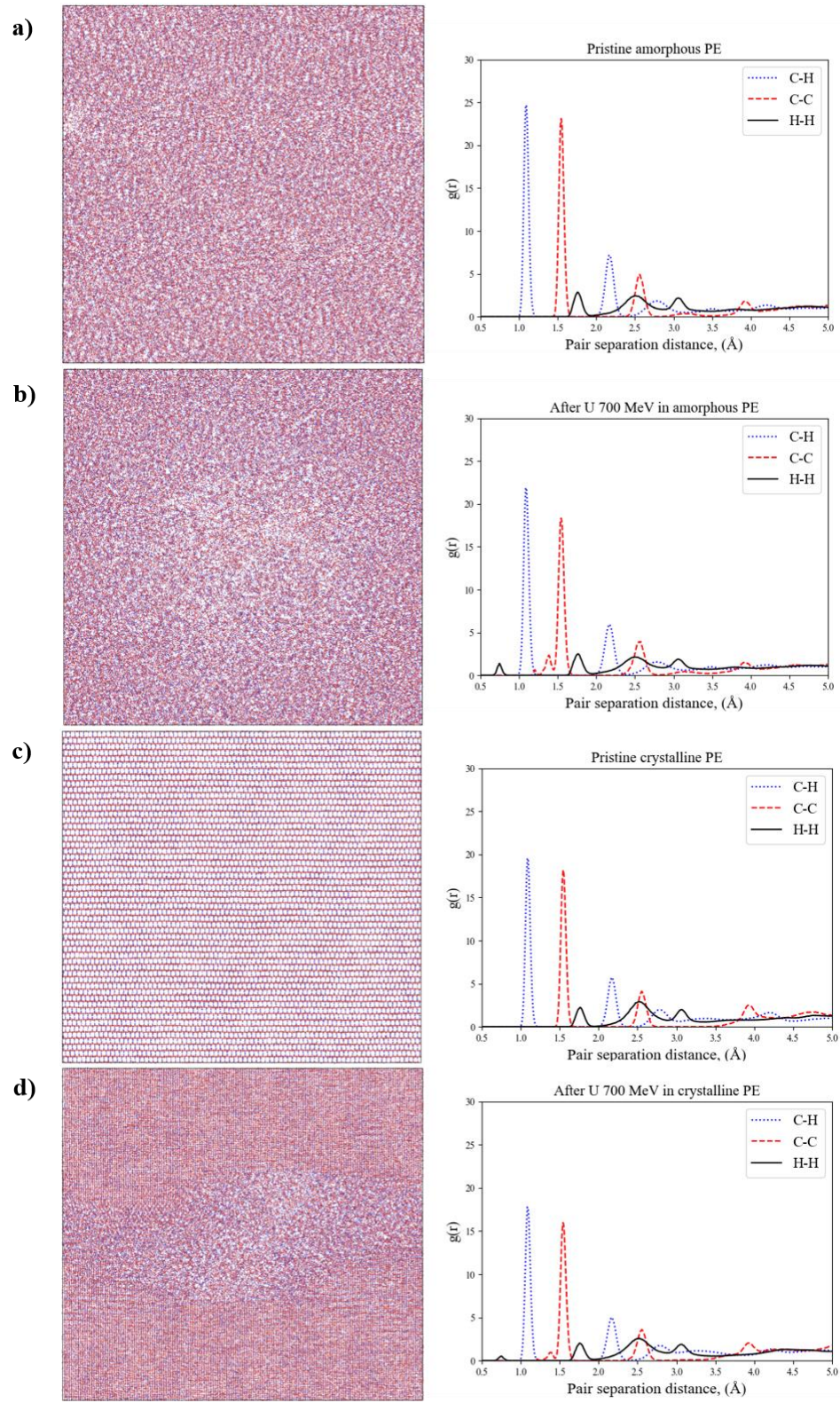

**Figure 3.** MD supercell snapshots (left panels) and RPDFs (right panels) prepared by OVITO [51]. (a) Pristine amorphous cell of PE; (b) amorphous one irradiated with U at the Bragg peak (700 MeV); (c) pristine crystalline PE cell; (d) crystalline one irradiated with uranium U at the Bragg peak.

Figure 4 presents the temporal evolution of the distribution of unsaturated carbon atoms in the amorphous and crystalline polyethylene supercells after the 1470 MeV Xe ion impact. We define an unsaturated carbon atom as an atom with dangling bonds (or/and having double-triple bonding) and thus having less than 4 neighbors belonging to it within the cut-off radius of interaction (2 Å for C-C and 1.8 Å for C-H interaction in AIREBO-M potential [53]). The spatial distribution of the initial damage (at 1 ps) is cylindrically symmetric in both amorphous and crystalline PE. The distribution of unsaturated carbons in the amorphous sample (Fig. 4a) remains circular, while relaxation of this damaged region in the crystalline supercell (Fig. 4b) later causes enhanced structure transformations along PE chains (see also three-dimensional distributions in Supplementary Materials, Figs. S1, S2).

Analysis of the number of unsaturated carbons shows that the increase in the number of bond breaks significantly slows down by about 25 picoseconds in both amorphous and crystalline cases. Further cooling of the track shows an expansion of the damaged area with these defects without a notable increase in their total number. Figure 4 (see also Figures S3-S6 in Supplementary Materials) shows that the number of unsaturated carbon atoms and their positions almost do not change after 250 ps in the amorphous sample and after 100 ps in the crystalline one. Cooling down to 350 K takes ~1 ns which is an order of magnitude longer than that in inorganic insulators [2].

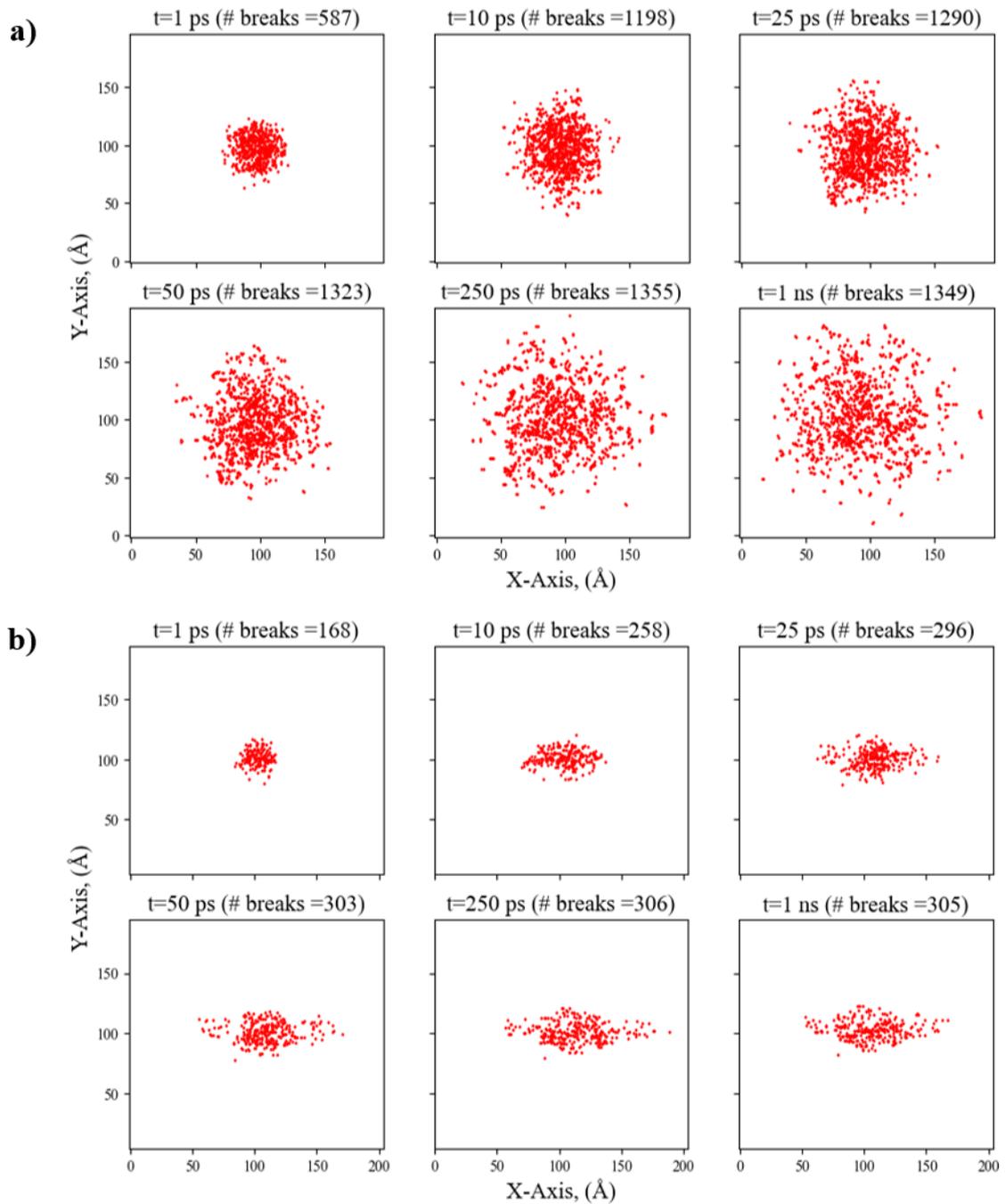

**Figure 4.** Spatio-temporal distribution of unsaturated carbons at different times after irradiation with 1470 MeV $^{129}$Xe ion in **(a)** amorphous and **(b)** crystalline PE. The projection of the cell perpendicular to the direction of the SHI trajectory is shown.

Figure 5 compares the regions of unsaturated carbons with the area of the decreased density and the experimentally measured track (observed as the region stained with a chemical colorant around the SHI's trajectory [52]).

As was mentioned above, the calculated damage in the amorphous supercell has a nearly circular shape due to the lack of specific chain directions and the radial energy deposition. The experimentally measured tracks in the amorphous samples irradiated with Xe ions also have a circular shape of an average diameter of 62.8 ± 20.4 Å [52]. In contrast, the damage in the crystalline supercell (Fig. 5b) has an elongated shape along the PE chains similar to that observed in the experiment with a width of 40-60 Å and a length of about 20 nm [52].

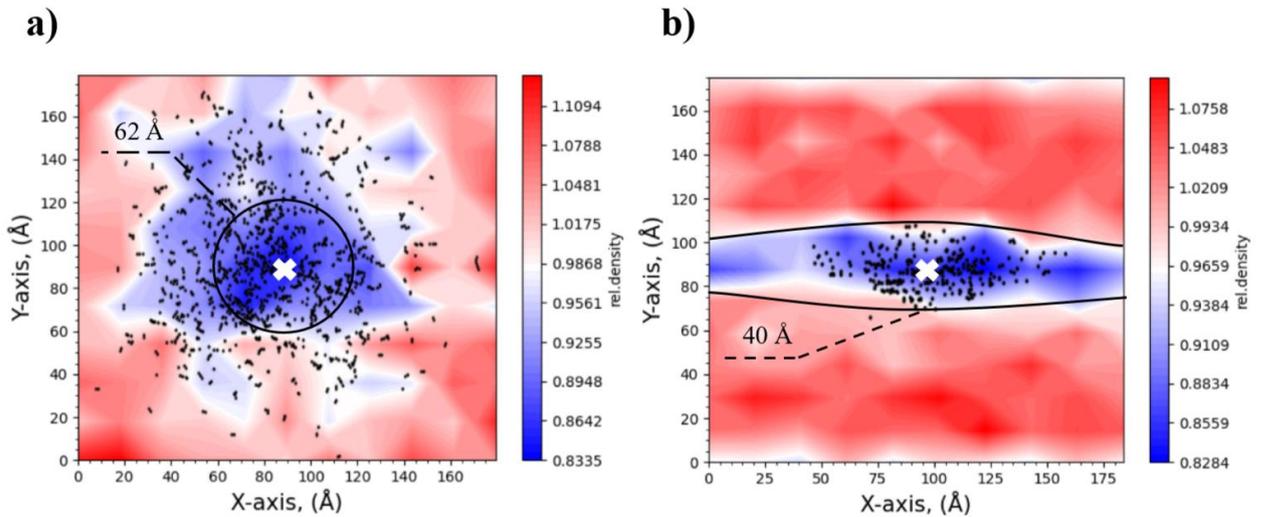

**Figure 5.** Calculated spatial distributions of unsaturated carbons and relative densities in **(a)** amorphous and **(b)** crystalline PE after 1470 MeV $^{129}$Xe impact. The black circle in **(a)** shows the average experimental track diameter (62.8 ± 20.4 Å) [52]; two black arcs in **(b)** present the minimal registered track width – 40 Å [52]. White crosses mark the SHI impact points.

In Figure 6 the track sizes calculated for U and Xe ions at 11.4 MeV/u irradiation are compared with the experimental data [52]. The track size decreases significantly with the atomic number of the SHI in our simulation, in agreement with the observation in [52]. In the case of Zn, the simulation shows no dangling bonds. Although the Zn ion track diameters are reported in [52] to be in the range of 24-58 Å, they are poorly visible in TEM (see Fig. 4 in [52]). In addition, the authors of [52] noted that the Zn tracks are unstable under electron beam irradiation in TEM and partially recover during observation, making a direct comparison with the simulation impossible. In [52], the authors noted that the exact mechanism of adhesion of molecules of the colorant (applied *post-mortem*) to PE in a damaged track is unknown. It was assumed that colorant molecules attach to the areas of high damage or a higher free volume, enabling visualization of the produced damage. There is a decrease in the material density in the proximity of the Zn trajectory in our simulation (see Figure S7 in Supplementary Materials). Our results suggest that the applied colorant can attach to sites of reduced density even in the absence of unsaturated carbons.

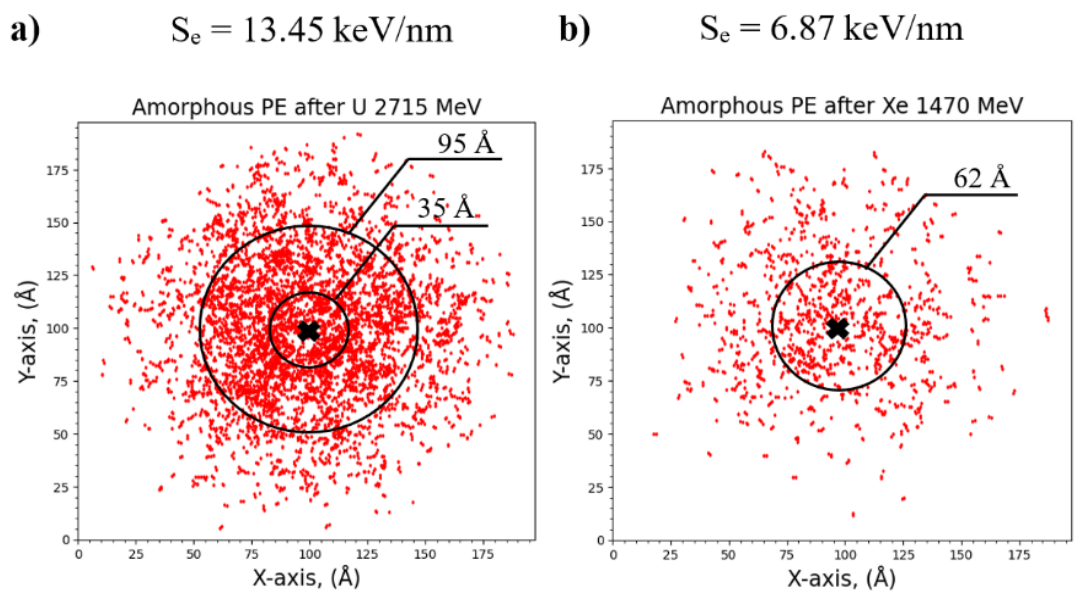

**Figure 6.** Unsaturated carbons after irradiation with the ions (a) U and (b) Xe, with the energy of 11.4 MeV/a.m.u. The solid circles indicate the lower and the upper diameters of the track, according to Ref. [52]. Crosses mark SHI's penetration points. The projection of the cell perpendicular to the direction of the SHI trajectory is shown.

More generally, a detailed quantitative comparison of the simulation with the experiment requires the development of models describing the chemical kinetics of colorant-PE interaction, which is beyond the scope of the present work.

## III.2. Polymer chain damage and fragment distributions

Figure 7 visualizes only intact chains highlighting the most damaged areas whose sizes and shapes coincide with those observed in the experiment.

**a)**                                                    **b)**

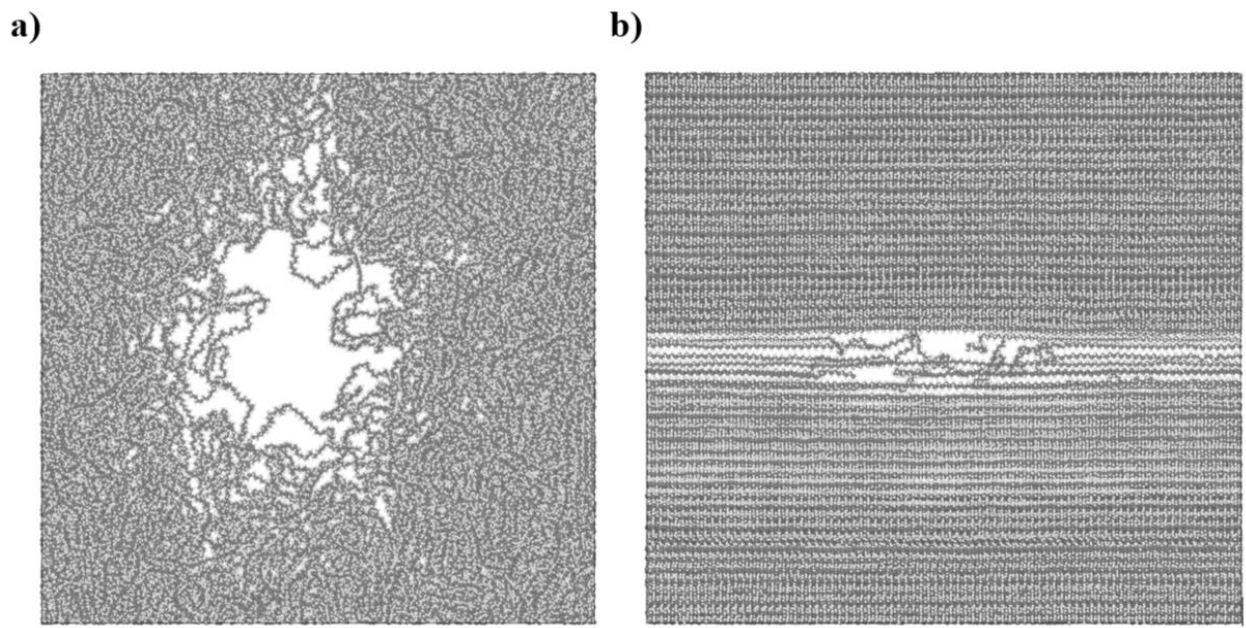

**Figure 7.** Supercells showing **(a)** undamaged chains in amorphous PE and **(b)** undamaged and longer cross-linked chains in crystalline PE irradiated with 1470 MeV [129]Xe.

Figure 8 illustrates the mass spectrum of fragments produced after the cooling. Mainly, hydrogen dimers and unbroken chains with a small contribution of various molecular forms including free radicals are created by SHI impact. In amorphous PE, fragments with masses between molecular hydrogen and intact chains are present (see inset in Fig. 8a). The mass distribution in crystalline PE differs from that in the amorphous one – fragments are compactly grouped near the two main peaks corresponding to molecular hydrogen and intact chains. Their relative quantity is much lower than in amorphous PE (see inset in Fig. 8b). Additionally, some of the fragments cross-linked during relaxation resulting in a small number of chains longer than the initial ones. 3D figures showing various molecular fragments can be found in Supplementary Materials (Figures S8, S9).

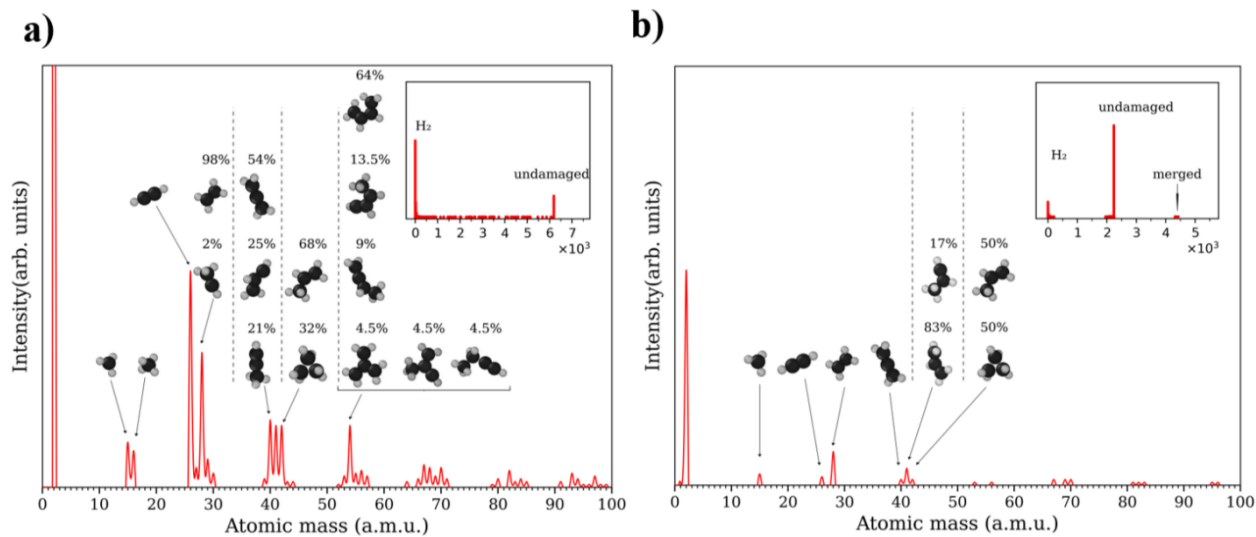

**Figure 8.** Mass distribution of fragments in a track in **a)** amorphous PE and **b)** crystalline lamella after the passage of Xe ion with energy 1470 MeV (the insets show the full range of masses). The percentages present depicted fragments' proportions corresponding to individual peaks; carbon and hydrogen atoms are shown in black and gray, respectively.

The spatial distributions of low-mass fragments (under 1000 a.m.u.) are similar to those of the unsaturated carbons and visible damage (cf. Figs.4,5,9,10). The distribution of small fragments is circular in the amorphous case and elongated in the crystal (Fig. 9). The same trend is seen for the fragments of higher masses (Fig. 10).

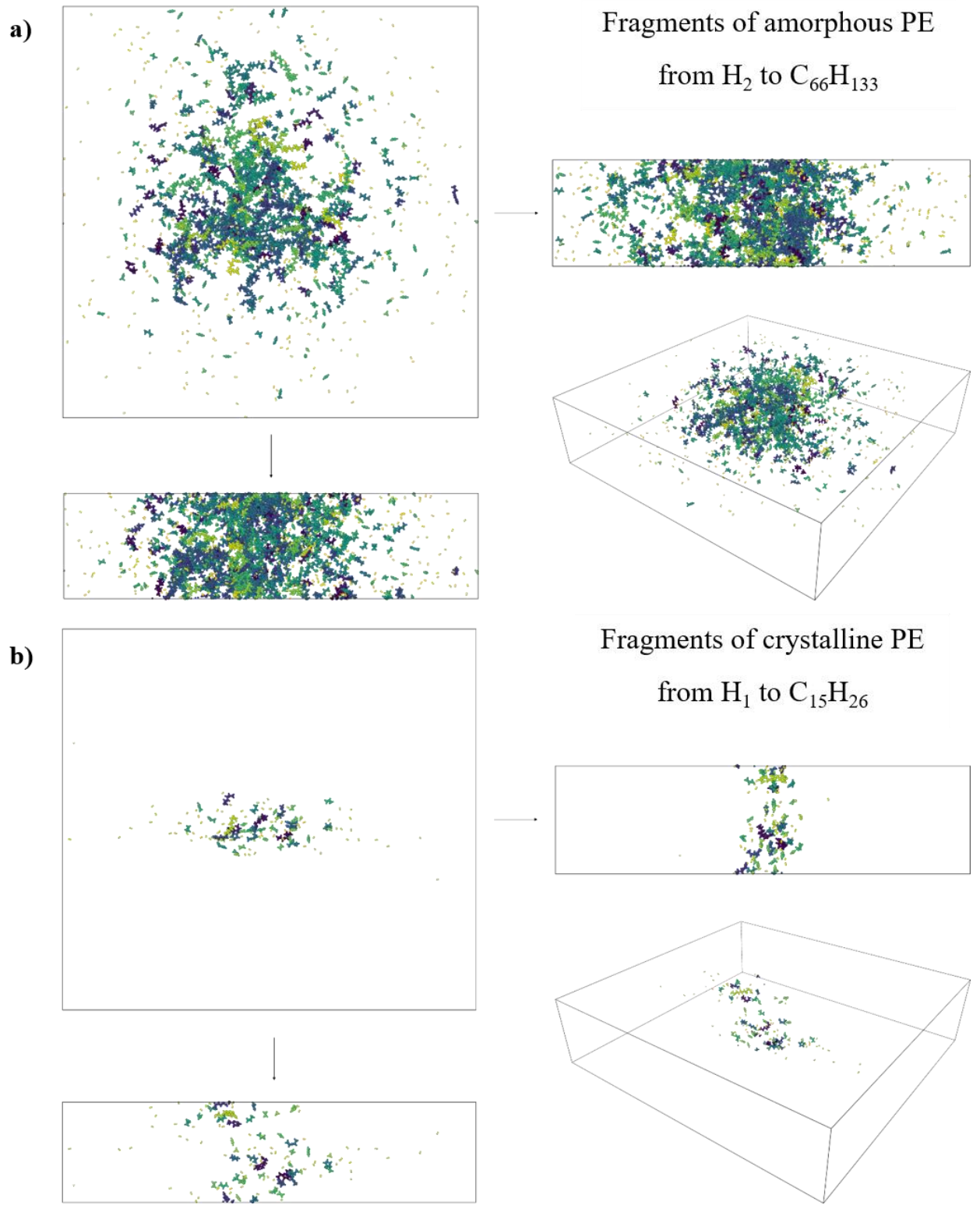

**Figure 9.** Distribution of fragments with masses < 1000 a.m.u. after 1470 MeV Xe ion passage in (**a**) amorphous and (**b**) crystalline PE. The darker color the heavier fragments.

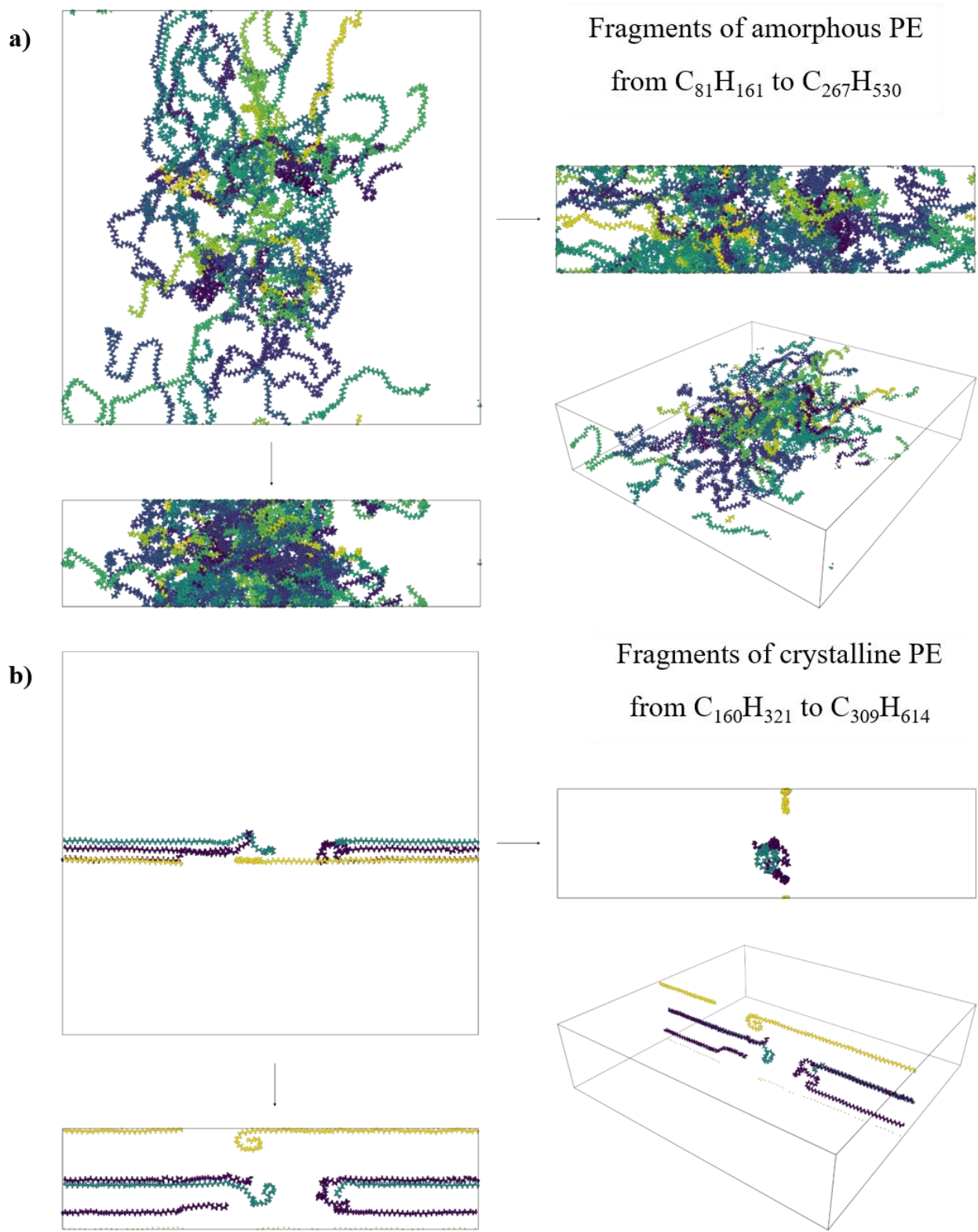

**Figure 10.** Distribution of fragments with masses from 1000 to 4000 a.m.u. after 1470 MeV Xe ion passage in (**a**) amorphous and (**b**) crystalline PE. The darker color the heavier fragments.

### III.3. Velocity effect in PE

Due to the peak-shaped Bragg curve of swift heavy ions energy losses, the same LET value can be achieved at two different SHI energies: on the left and the right shoulders of the curve, see Figure 11. These energies can differ by an order of magnitude, resulting in very different spectra of fast electrons excited by the ions [27]. The different spectra result in the different radial distributions of the energy deposited to the target lattice causing different damage. This velocity effect is well-known in inorganic materials [1,3] and has recently been shown in polyethylene terephthalate [10].

We chose two energies 100 MeV and 5500 MeV for $^{238}$U ions, producing the same LET of 10.9 keV/nm (70% of the Bragg peak value). It appears that the slower SHI causes a higher density of bond breaks, confirming the existence of the velocity effect in amorphous and crystalline PE samples (see Fig. 11).

**a)**

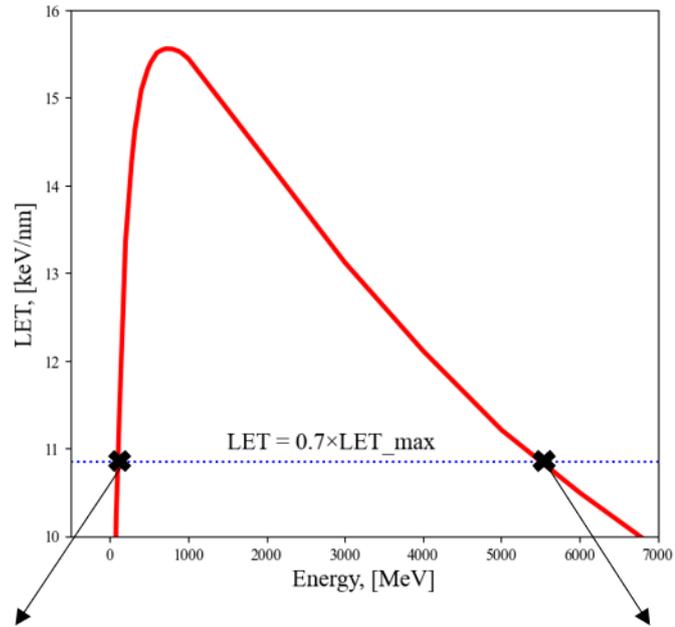

**b)** Crystalline PE after U 100 MeV (# breaks =4648)

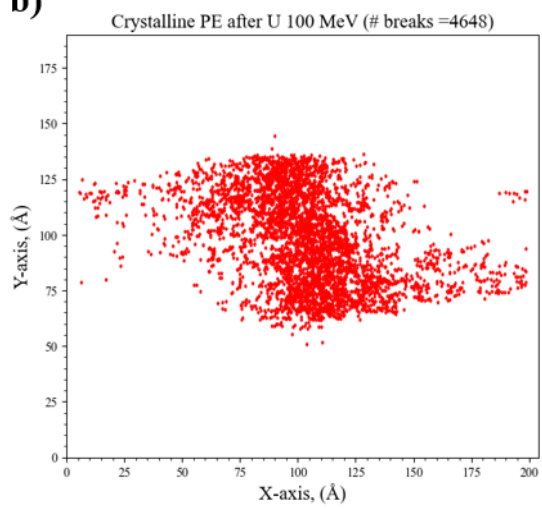

**c)** Crystalline PE after U 5500 MeV (# breaks =1904)

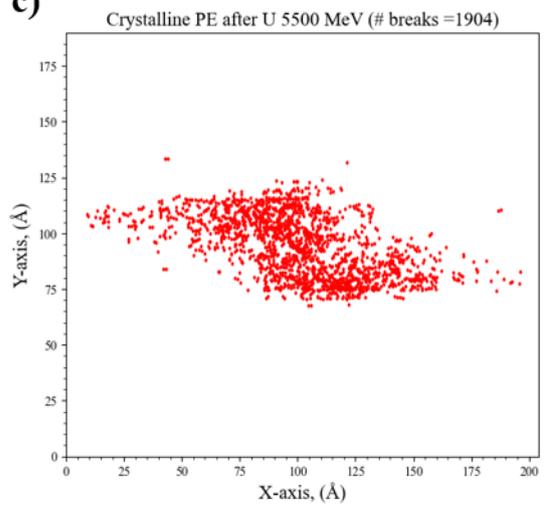

**d)** Amorphous PE after U 100 MeV (# breaks =7405)

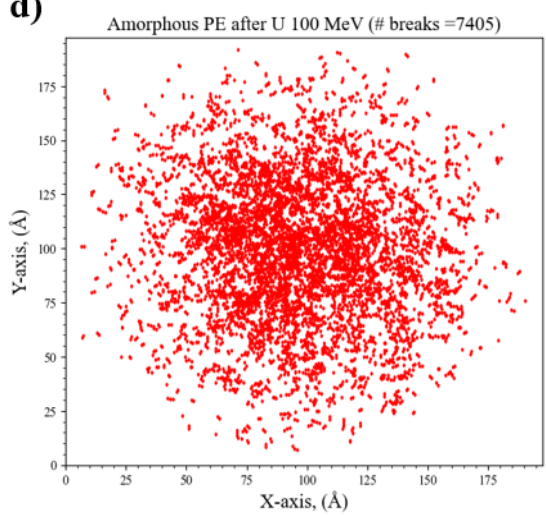

**e)** Amorphous PE after U 5500 MeV (# breaks =3889)

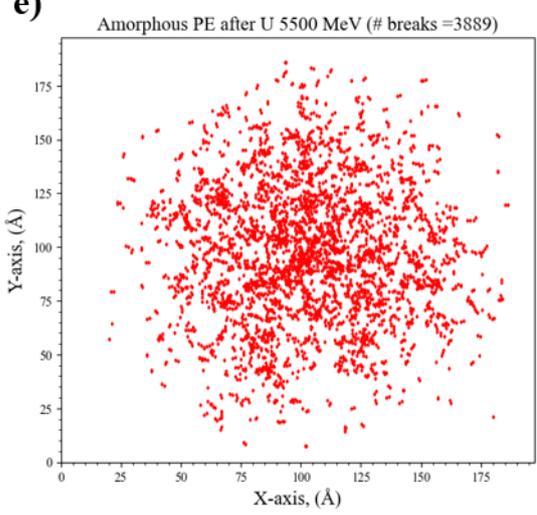

**Figure 11. (a)** LET curve for $^{238}$U calculated with TREKIS-3. Energies 100 MeV and 5500 MeV producing the same LET = 11 keV/nm are marked with the horizontal dotted line. (**b-e**) Snapshots of unsaturated carbon atoms after simulation with TREKIS+MD of $^{238}$U passage in crystalline PE specimen with energies 100 and 5500 MeV (**b, c**) and $^{238}$U passage in amorphous PE with energies 100 and 5500 MeV (**d, e**).

The skew of the spatial distribution of the unsaturated carbons to the left is evident in Fig. 11 b, c. The simulations showed that the distribution of the unsaturated atoms can arbitrarily lean to the left, to the right, stay straight or shift in the supercell as a whole due to fluctuations in the initial conditions in MD calculations (see Figures S10, S11 in Supplementary Materials).

## Conclusions

Monte Carlo code TREKIS-3 combined with the molecular dynamics code LAMMPS enabled us to describe excitation and structural changes in amorphous and crystalline polyethylene irradiated with swift heavy ions without *a posteriori* fitting parameters. The evolution of the atomic structure, showing damaged chemical bonds, is followed up to the time when no further structural changes occur (~1 ns). In amorphous polyethylene, initial bond ruptures form a circular dense core that expands radially during the cooling to form finally a nearly circular track by ~250 ps. In contrast, in the crystalline sample, an elliptical region of damaged bonds stretched along the polymer chains evolves from the initial cylindrical excitation over the timescale of tens of picoseconds. Our simulations agree with the experimentally measured track sizes and also reproduce the difference in the shapes in the crystalline and amorphous samples, validating the model.

The obtained radial pair distribution functions indicate significant damage in the vicinity of the SHI track. Analysis of the mass spectra of the simulated sample shows the presence of a large amount of molecular hydrogen and low-mass fragments in the track. No voids were found in tracks in bulk PE, in agreement with the experiment [52].

We also demonstrated the velocity effect in PE: ions with the same linear energy transfer but different velocities (the left and the right shoulders of the Bragg curve) cause different damage in the material. The lower velocity ion causes more damage over a larger area. This effect has been shown experimentally for PET polymer [10].

The results were obtained without fitting parameters in simulations, demonstrating the applicability of a combined model to polymers tracing both, SHI-induced electronic and atomic kinetics.

## Acknowledgments

The authors are grateful to Michael V. Sorokin for helpful discussions. PB, SG, RV, and AEV gratefully acknowledge financial support from the Russian Science Foundation (grant No. 22-22-00676). NM thanks the financial support from the Czech Ministry of Education, Youth, and Sports (grants No. LTT17015, LM2018114, and No. EF16_013/0001552). This work has been carried out using computing resources of the federal collective usage center Complex for Simulation and Data Processing for Mega-science Facilities at NRC "Kurchatov Institute", " http://ckp.nrcki.ru".